\documentclass[journal]{new-aiaa}

\usepackage[utf8]{inputenc} %
\usepackage[T1]{fontenc} %
\usepackage[english]{babel} %
\usepackage{amsmath, wasysym}
\usepackage{graphicx,setspace,parskip,bm} %
\usepackage{mathtools} %
\usepackage{tabularx,multicol,multirow,threeparttablex} %
\usepackage{tablefootnote} %
\usepackage{footnpag,marginnote} %
\usepackage{booktabs} %
\usepackage{caption} %
\usepackage{subcaption} %
\usepackage{xcolor} %
\usepackage{url} %
\usepackage{siunitx} %
\usepackage{comment} %
\usepackage{lipsum} %
\usepackage{appendix} %
\usepackage[mathlines]{lineno} %
\usepackage[nonumberlist,acronym]{glossaries} %

\graphicspath{{figs/}}

\hypersetup{
    hidelinks
}

\newtheorem{definition}{Definition}
\newtheorem{problem}{Problem}

\DeclareSIUnit{\AU}{AU}
\DeclareSIUnit{\day}{day}
\DeclareSIUnit{\days}{days}
\DeclareSIUnit{\deg}{deg}
\DeclareSIUnit{\year}{year}
\DeclareSIUnit{\years}{years}
\DeclareSIUnit{\LU}{LU}
\DeclareSIUnit{\TU}{TU}
\DeclareSIUnit{\VU}{VU}
\DeclareSIUnit{\MU}{MU}
\DeclareSIUnit{\ACU}{ACU}

\newcommand{\transpose}{\top}
\newcommand{\Eq}[1]{Eq.~\eqref{#1}}

\newcommand{\Fig}[1]{Fig.~\ref{#1}}
\newcommand{\Figs}[1]{Figs.~\ref{#1}}
\newcommand{\Tab}[1]{Table~\ref{#1}}

\newcommand{\Sec}[1]{Section~\ref{#1}}

\newcommand{\ie}{i.\,e.,~}

\newcommand{\Numtxt}[1]{$\mathrm{#1}$}
\newcommand{\lastdate}{Sep 1, 2023}

\newcommand{\WSBset}[3]{$\mathcal{#1}_{#2}^{#3}$}
\newcommand{\Wset}[1]{\WSBset{W}{#1}{}}
\newcommand{\Xset}[1]{\WSBset{X}{#1}{}}
\newcommand{\Kset}[1]{\WSBset{K}{#1}{}}
\newcommand{\Dset}[1]{\WSBset{D}{#1}{}}
\newcommand{\Mset}[1]{\WSBset{M}{#1}{}}
\newcommand{\Cset}[1]{\WSBset{C}{-1}{#1}}
\newcommand{\corridor}[1]{\WSBset{\check{B}}{-1}{#1}}
\newcommand{\Scorr}[1]{\WSBset{\check{S}}{-1}{#1}}
\newcommand{\Icorr}[1]{\WSBset{\check{I}}{-1}{#1}}
\newcommand{\Ecorr}[1]{\WSBset{\check{E}}{-1}{#1}}
\newcommand{\Envcorr}[1]{\WSBset{\partial \check{S}}{-1}{#1}}

\newcommand{\Sun}{\odot}

\newcommand{\textfnc}[1]{{\fontfamily{pcr}\selectfont #1}}

\newacronym{3PBVP}{3PBVP}{three-point boundary value problem}
\newacronym{ABC}{ABC}{autonomous ballistic capture}
\newacronym{ABM}{ABM}{Adams--Bashforth--Moulton}
\newacronym{ASI}{ASI}{Agenzia Spaziale Italiana}
\newacronym{BC}{BC}{ballistic capture}
\newacronym{BCC}{BCC}{ballistic capture corridor}
\newacronym{BER4BP}{BER4BP}{bi-elliptic restricted 4-body problem}
\newacronym{BME}{BME}{body mean equator of date frame}
\newacronym{CGST}{CGST}{Cauchy--Green strain tensor}
\newacronym{CK}{CK}{orientation or C-matrix kernel}
\newacronym{CNES}{CNES}{Centre National d'\'Etudes Spatiales}
\newacronym{CR3BP}{CR3BP}{circular restricted three-body problem}
\newacronym{DA}{DA}{differential algebra}
\newacronym{DART}{DART}{Deep-space Astrodynamics Research \& Technology}
\newacronym{DOPRI8}{DOPRI8}{Dormand--Prince 8th-order embedded Runge--Kutta}
\newacronym{DRO}{DRO}{distant retrograde orbit}
\newacronym{ECSS}{ECSS}{European Cooperation for Space Standardization}
\newacronym{EOM}{EoM}{Equations of motion}
\newacronym{ER3BP}{ER3BP}{elliptic restricted three-body problem}
\newacronym{ERC}{ERC}{European Research Council}
\newacronym{ESA}{ESA}{European Space Agency}
\newacronym{ESH}{ESH}{EXTREMA simulation hub}
\newacronym{FK}{FK}{frame kernel}
\newacronym{FLI}{FLI}{fast Lyapunov indicator}
\newacronym{FTLE}{FTLE}{finite-time Lyapunov exponent}
\newacronym{GMAT}{GMAT}{General Mission Analysis Tool}
\newacronym{GNC}{GNC}{guidance, navigation, and control}
\newacronym{GRATIS}{GRATIS}{GRavity TIdal Slide}
\newacronym{GSTP}{GSTP}{General Support Technology Programme}
\newacronym{IC}{IC}{initial condition}
\newacronym{ICRF}{ICRF}{International Celestial Reference Frame}
\newacronym{IK}{IK}{instrument kernel}
\newacronym{JAXA}{JAXA}{Japan Aerospace Exploration Agency}
\newacronym{JPL}{JPL}{Jet Propulsion Laboratory}
\newacronym{LCS}{LCS}{Lagrangian coherent structure}
\newacronym{LD}{LD}{Lagrangian descriptor}
\newacronym{LHS}{LHS}{Latin hypercube sampling}
\newacronym{LSK}{LSK}{leap seconds kernel}
\newacronym{LUMIO}{LUMIO}{Lunar Meteoroid Impacts Observer}
\newacronym{MC}{MC}{Monte Carlo}
\newacronym{MEGNO}{MEGNO}{mean exponential growth factor of nearby orbits}
\newacronym{MMX}{MMX}{Martian Moons eXploration}
\newacronym{NAIF}{NAIF}{Navigation and Ancillary Information Facility}
\newacronym{NASA}{NASA}{National Aeronautics and Space Administration}
\newacronym{NEA}{NEA}{near-Earth asteroid}
\newacronym{NEO}{NEO}{near-Earth object}
\newacronym{NSG}{NSG}{non-spherical gravity}
\newacronym{ORQ}{ORQ}{operational research question}
\newacronym{PBER4BP}{PBER4BP}{planar bi-elliptic restricted four-body problem}
\newacronym{PCC}{PCC}{polynomial chaos coefficient}
\newacronym{PCE}{PCE}{polynomial chaos expansion}
\newacronym{PCK}{PCK}{planetary constant kernel}
\newacronym{PDE}{PDE}{partial differential equation}
\newacronym{PECE}{PECE}{predictor-corrector}
\newacronym{PhD}{PhD}{Philosophi\ae \ Doctor}
\newacronym{RAAN}{RAAN}{right ascension of the ascending node}
\newacronym{RHS}{RHS}{right-hand side}
\newacronym{RK}{RK}{Runge--Kutta}
\newacronym{RO}{RO}{research objective}
\newacronym{RQ}{RQ}{research question}
\newacronym{RSS}{RSS}{root sum square}
\newacronym{RT}{R\&T}{Research and Technology}
\newacronym{RTN}{RTN@$t_{i}$}{radial-tangential-normal of date frame}
\newacronym{RPF}{RPF}{roto-pulsating frame}
\newacronym{SCLK}{SCLK}{spacecraft clock kernel}
\newacronym{SOI}{SOI}{sphere of influence}
\newacronym{SPK}{SPK}{spacecraft ephemeris kernel}
\newacronym{SRP}{SRP}{solar radiation pressure}
\newacronym{SSTO}{SSTO}{Sun-synchronous terminator orbit}
\newacronym{STM}{STM}{state transition matrix}
\newacronym{UOM}{uom}{unit of measurement}
\newacronym{VSVO}{VSVO}{variable-step, variable-order}
\newacronym{VV}{V\&V}{verification and validation}
\newacronym{WSB}{WSB}{weak stability boundary}

\title{Synthesis of Ballistic Capture Corridors at Mars via Polynomial Chaos Expansion}

\author{M. Liotta\footnote{MSc graduate, martina.liotta@mail.polimi.it.}, G. Merisio\footnote{Post Doctoral Research Fellow, Department of Aerospace Science and Technology, gianmario.merisio@polimi.it.}, C. Giordano\footnote{Post Doctoral Research Fellow, Department of Aerospace Science and Technology, carmine.giordano@polimi.it; AIAA Member.} and F. Topputo\footnote{Professor, Department of Aerospace Science and Technology, francesco.topputo@polimi.it; AIAA Senior Member.}.}
\affil{Politecnico di Milano, Via La Masa 34, Milano, 20156, Italy}

\begin{document}

\maketitle

\section{Introduction}

The space sector is experiencing a flourishing growth. Evidence is mounting that the near future will be characterized by a large amount of deep-space missions \cite{poghosyan2017cubesat,bandyopadhyay2016review,kalita2017network,hein2018exploring}. In the last decade, CubeSats have granted affordable access to space due to their reduced manufacturing costs compared to traditional missions. Nowadays, most miniaturized spacecraft have thus far been deployed into near-Earth orbits, but soon a multitude of interplanetary CubeSats will be employed for deep-space missions as well \cite{didomenico2022erc}. Nevertheless, the current paradigm for deep-space missions strongly relies on ground-based operations \cite{turan2022autonomous}. Although reliable, this approach will rapidly cause saturation of ground slots, thereby hampering the current momentum in space exploration. At the actual pace, human-in-the-loop, flight-related operations for deep-space missions will soon become unsustainable.

Self-driving spacecraft challenge the current paradigm under which spacecraft are piloted in interplanetary space. They are intended as machines capable of traveling in deep space and autonomously reaching their destination. In EXTREMA (short for \emph{Engineering Extremely Rare Events in Astrodynamics for Deep-Space Missions in Autonomy}) \cite{didomenico2022erc,andreis2022onboard}, these systems are used to engineer \gls{BC} \cite{hyeraci2010method,luo2014constructing,deitos2018survey}, thereby proving the effectiveness of autonomy in a complex scenario. The \gls*{BC} mechanism allows capture about a planet exploiting the natural dynamics, thus without requiring maneuvers \cite{topputo2015earth,belbruno1993sun,belbruno2000calculation,quinci2023qualitative}. At the expense of longer transfer times, \gls*{BC} orbits are cheaper, safer, and more versatile from the operational perspective than Keplerian solutions \cite{topputo2015earth}. Furthermore, \gls{BC} is a desirable solution for limited-control platforms, which cannot afford to enter into orbits about a planet due to a lack of significant control authority. The key is to accomplish low-thrust orbits culminating in \gls{BC}. For this, a bundle of \gls*{BC} orbits named \gls*{BCC} can be targeted far away from a planet \cite{merisio2023engineering,morelli2022convex,aguiar2018technique}. Mars is chosen in this work due to its relevance in the long-term exploration \cite{didomenico2022erc}.

\gls*{BC} is an extremely rare event, thus massive numerical simulations are required to find the specific conditions supporting capture. On average, only 1 out of \num{10000} conditions explored by algorithms grants capture \cite{luo2015analysis}. To achieve \gls*{BC} at Mars without any a priori instruction, an inexpensive and accurate method to construct \gls*{BCC} directly on board is required. Therefore, granting spacecraft the capability to manipulate stable sets in order to compute autonomously \gls{BCC} is crucial \cite{merisio2023engineering}. The goal of the paper is to numerically synthesize a corridor exploiting the \gls*{PCE} technique, thereby applying a suited uncertainty propagation technique to \gls*{BC} orbit propagation. In the literature, \gls*{PCE} was introduced and exploited for uncertainty quantification \cite{giordano2021analysis,jones2013nonlinear}. \gls*{PCE} can be used as an effective interpolation method, avoiding a priori definition of interpolant functions but selecting them automatically starting from the input samples, so that they possess spectral convergence with respect to the input variables. Furthermore, proving it to be successful in propagating all-in-once a bundle of trajectories in a deterministic setting opens the door to a wider use for \gls*{PCE} in data-driven approaches \cite{pugliatti2023image}. The proposed approach is validated against \gls*{MC} simulation. The heavy computational loads derived by multiple point-wise propagations of \gls*{BC} orbits when performing guidance and control tasks during the low-thrust interplanetary cruise are unburdened by the devised methodology. Broadly, this will facilitate the paradigm shift towards autonomy, thereby favoring the reduction of mission load on ground stations by decreasing the demand foreseen in the next years.

The remainder of the paper is organized as follows. In \Sec{sec:background}, the background is introduced. Then, the proposed methodology follows in \Sec{sec:method}. Results are presented in \Sec{sec:results}. Eventually, conclusions are drawn in \Sec{sec:conclusion}.

\section{Background} \label{sec:background}

\subsection{Dynamical model}\label{sse:dyn_mod}

Following the nomenclature in \cite{luo2014constructing}, a \textit{target} and a \textit{primary} are defined. The target is the body around which \gls*{BC} is studied. The primary is the main body around which the target revolves. Target and primary masses are $m_{t}$ and $m_{p}$, respectively. The mass ratio of the system is $\mu = m_{t} / (m_{t} + m_{p})$. This work focuses on \gls*{BC} having Mars as target and the Sun as primary. Reference frames used in this work are the J2000 and \acrshort{RTN} \cite{caleb2022stable}.

The precise states of the Sun and the major planets are retrieved from the \gls*{JPL}'s planetary ephemerides \textfnc{de440s.bsp}\footnote{Data publicly available at: \url{https://naif.jpl.nasa.gov/pub/naif/generic_kernels/spk/planets/de440s.bsp} [retrieved \lastdate].} (or DE440s) \cite{park2021jpl}. Additionally, the ephemerides \textfnc{mar097.bsp} of Mars (the target) and its moons are employed\footnote{Data publicly available at: \href{https://naif.jpl.nasa.gov/pub/naif/generic_kernels/spk/satellites/mars097.bsp}{\nolinkurl{~/spk/satellites/mar097.bsp}} [retrieved \lastdate].}. The following generic \gls*{LSK} and \gls*{PCK} are used: \textfnc{naif0012.tls}, \textfnc{pck00010.tpc}, and \textfnc{gm\_de440.tpc}\footnote{Data publicly available at: \href{https://naif.jpl.nasa.gov/pub/naif/generic_kernels/lsk/naif0012.tls}{\nolinkurl{~/lsk/naif0012.tls}}, \href{https://naif.jpl.nasa.gov/pub/naif/generic_kernels/pck/pck00010.tpc}{\nolinkurl{~/pck/pck00010.tpc}}, and \href{https://naif.jpl.nasa.gov/pub/naif/generic_kernels/pck/gm_de440.tpc}{\nolinkurl{~/pck/gm_de440.tpc}} [retrieved \lastdate].}.

The \gls*{EOM} of the restricted $n$-body problem are considered. The gravitational attractions of the Sun, Mercury, Venus, Earth--Moon (B\footnote{Here B stands for barycenter.}), Mars (central body), Jupiter (B), Saturn (B), Uranus (B), and Neptune (B) are taken into account. Additionally, \gls*{SRP}, Mars' \gls*{NSG}, and relativistic corrections \cite{huang1990relativistic} are considered. \Tab{tab:sc-specs} collects the assumed spacecraft parameters needed to evaluate the \gls*{SRP} perturbation. They are compatible with those of a 12U deep-space CubeSat \cite{topputo2021envelop}. Terms of the infinite series modeling \gls*{NSG} are considered up to degree $ n_{\mathrm{deg}} = \SI{20}{} $ and order $ n_{\mathrm{ord}} = \SI{20}{} $ \cite{aguiar2018technique}. The coefficients to evaluate the \gls*{NSG} perturbation are retrieved from the MRO120F gravity field model of Mars. Data are publicly available in the file \textfnc{jgmro\_120f\_sha.tab}, archived in the Geosciences Node of NASA’s Planetary Data System\footnote{Data publicly available at: \url{https://pds-geosciences.wustl.edu/mro/mro-m-rss-5-sdp-v1/mrors_1xxx/data/shadr/} [retrieved \lastdate].}. Far from Mars, when in heliocentric motion, the \gls*{NSG} perturbation is neglected. \gls*{EOM} are integrated in the J2000 inertial frame.

\begin{table}[tbp]
    \centering
    \caption[Spacecraft parameters for SRP evaluation]{Spacecraft parameters for SRP evaluation \cite{topputo2021envelop}.}
    \begin{tabular}{l c c}
        \toprule
        \textbf{Parameter} & \textbf{Unit} &\textbf{Value} \\
        \midrule
        Mass--SRP area ratio $ m / A $ & \si{\kilogram\per\meter\squared} & \SI{75}{} \\
        Coefficient of reflectivity $ C_{r} $ & - & \SI{1.3}{} \\
        \bottomrule
    \end{tabular}
    \label{tab:sc-specs}
\end{table}  

The \gls*{EOM} in a non-rotating Mars-centered reference frame are \cite{luo2014constructing,deitos2019high,huang1990relativistic}
\begin{equation}
    \begin{split}
	\ddot{\mathbf{r}} = & - \frac{\mu_{t}}{r^{3}} \mathbf{r} 
	- \sum\limits_{i \in \mathbb{P}} \mu_{i} \left( \frac{\mathbf{r}_{i}}{r_{i}^{3}} + \frac{\mathbf{r} - \mathbf{r}_{i}}{\left\lVert \mathbf{r} - \mathbf{r}_{i} \right\rVert^{3}} \right)
	+ \frac{Q A}{m} \frac{\mathbf{r} - \mathbf{r}_{\Sun}}{\left\lVert \mathbf{r} - \mathbf{r}_{\Sun} \right\rVert^{3}}
	- \mathcal{R} \frac{\mu_{t}}{r^{2}} \left( \Lambda  \frac{\mathcal{R}^\top \mathbf{r}}{r} - \left[ J^{\transpose} \ K^{\transpose} \ H^{\transpose} \right]^{\transpose} \right) + \\
	& + \frac{\mu_{t}}{c^{2}r^{3}} \left[ \left( 4\frac{\mu_{t}}{r} - v^{2} \right) \mathbf{r} + 4 \left( \mathbf{r} \cdot \dot{\mathbf{r}} \right) \dot{\mathbf{r}} \right]
	+ 2 \left( \mathbf{\Omega} \times \dot{\mathbf{r}} \right)
	+ 2 \frac{\mu_{t}}{c^{2}r^{3}} \left[ \frac{3}{r^{2}} \left( \mathbf{r} \times \dot{\mathbf{r}} \right) \left( \mathbf{r} \cdot \mathbf{J} \right) + \left( \dot{\mathbf{r}} \times \mathbf{J} \right) \right]
	\label{eq:eom}
    \end{split}
\end{equation}
where $ \mu_{t} $ is the gravitational parameter of the target body (\ie Mars in this work); $ \mathbf{r} $ and $ \dot{\mathbf{r}} = \mathbf{v} $ are the position and velocity vectors of the spacecraft with respect to the target, respectively, being $ r $ and $ v $ their magnitudes; $ \mathbb{P} $ is a set of $ n - 2 $ indexes (where $n$ concerns the $n$-body problem) each one referring to the perturbing bodies; $ \mu_{i} $ and $ \mathbf{r}_{i} $ are the gravitational parameter and position vector of the $i$-th body with respect to the target, respectively; $ A $ is the Sun-projected area on the spacecraft for \gls*{SRP} evaluation; $ m $ is the spacecraft mass; $ \mathbf{r}_{\Sun} $ is the position vector of the Sun with respect to the target; $ \mathcal{R} $ is the time-dependent matrix transforming vector components from the Mars-fixed frame to the non-rotating frame in which the \gls*{EOM} are written; $ \Lambda $, $ J $, $ K $, and $ H $ are defined as in \cite{gottlieb1993fast}; $ c = \SI{299792458}{\meter\per\second} $ (from SPICE \cite{acton1996ancillary,acton2018look}) is the speed of light in vacuum; $ \mathbf{J} $ is the rotating central body's angular momentum per unit mass in the J2000 frame. Then, $ Q = L C_{r} / \left( 4 \pi c \right) $ where $ C_{r} $ is the spacecraft coefficient of reflectivity, and $ L = S_{\Sun} 4 \pi d_{\mathrm{AU}}^{2} $ is the luminosity of the Sun. The latter is computed from the solar constant\footnote{\url{https://extapps.ksc.nasa.gov/Reliability/Documents/Preferred_Practices/2301.pdf} [last accessed \lastdate].} $ S_{\Sun} = \SI{1367.5}{\watt\per\meter\squared} $  evaluated at $ d_{\mathrm{AU}} = \SI{1}{\AU} $. Lastly, $ \mathbf{\Omega} = \tfrac{3}{2} \dot{\mathbf{r}}_{\Sun / t} \times \left( - \mu_{\Sun} \mathbf{r}_{\Sun / t} \right) / \left( c^{2} r_{\Sun / t}^{3} \right) $ where $ \mu_{\Sun} $ is the gravitational parameter of the Sun; $ \mathbf{r}_{\Sun / t} $ and $ \dot{\mathbf{r}}_{\Sun / t} = \mathbf{v}_{\Sun / t}$ are the position and velocity vectors, respectively, of the target body with respect to the Sun, being $ r_{\Sun / t} $ and $ \dot{r}_{\Sun / t} = v_{\Sun / t} $ their magnitudes.

The \gls*{EOM} in \Eq{eq:eom} are integrated with the \gls*{GRATIS} tool \cite{topputo2018trajectory} in their nondimensional form to avoid ill-conditioning (see normalization units in \Tab{tab:units}) \cite{luo2014constructing}. Numerical integration is carried out with the DOPRI8 propagation scheme \cite{montenbruck2000satellite,prince1981high}. The dynamics are propagated with relative and absolute tolerances set to \Numtxt{10^{-13}} \cite{luo2014constructing}.

\begin{table}[t]
    \centering
    \begin{threeparttable}
        \caption[Nondimensionalization units]{Nondimensionalization units.}
        \begin{tabular}{llll}
            \toprule
            \textbf{Unit} & \textbf{Symbol} & \textbf{Value} & \textbf{Comment} \\
            \midrule
            Gravity parameter & $ \mathrm{MU} $ & \SI{42828.376}{\kilo\meter\cubed\per\second\squared} & Mars' gravity parameter $ \mu_{t} $ \\
            Length & $ \mathrm{LU} $ & \SI{3396.0000}{\kilo\meter} & Mars' radius $ R_{\mars} $ \\ 
            Time\tnote{\dag} & $ \mathrm{TU} $ & \SI{956.28142}{\second} & $ (\mathrm{LU}^3 / \mathrm{MU})^{0\!.5} $ \\
            Velocity & $ \mathrm{VU} $ & \SI{3.5512558}{\kilo\meter\per\second} & $ \mathrm{LU} / \mathrm{TU} $ \\
            \bottomrule
        \end{tabular}
        \label{tab:units}
        \begin{tablenotes}
            \scriptsize
            \item[\tnote{\dag}] Time unit chosen such that the nondimensional period of a circular orbit of radius \si{\LU} equals $ \num{2} \pi $.
        \end{tablenotes}
    \end{threeparttable}
\end{table}

\subsection{\Acrlongpl*{BCC}} \label{sec:bcc}
\Gls*{BC} orbits are characterized by \glspl*{IC} escaping the target when integrated backward and performing $n$ revolutions about it when propagated forward, neither impacting nor escaping the target. In forward time, particles flying on \gls*{BC} orbits approach the target coming from outside its sphere of influence and remain temporarily captured about it. After a certain time, the particle escapes if an energy dissipation mechanism does not take place to make the capture permanent. To dissipate energy either a breaking maneuver or the target atmosphere (if available) could be used \cite{luo2021mars,giordano2022aeroballistic}. Effects on \gls*{BC} by gravitational attractions of many bodies besides the primaries and \gls*{SRP} have been investigated in previous works \cite{merisio2021characterization,aguiar2018technique,luo2017capability,luo2015analysis}.

\begin{figure}[t]
    \centering
    \includegraphics[width=\linewidth]{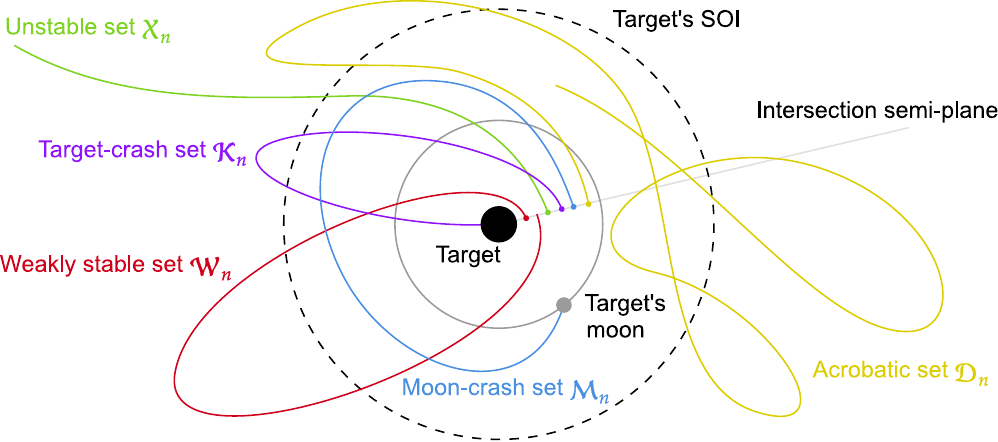}
    \caption{Illustration of particle stability. Weakly stable set \Wset{n} in red, unstable set \Xset{n} in green, target-crash set \Kset{n} in purple, moon-crash set \Mset{n} in blue, and acrobatic set \Dset{n} in yellow. Example with $ n = \num{1} $. The points on the intersection semi-plane are the ICs.}
    \label{fig:particle_stability}
\end{figure}

A particle stability is inferred using a plane in the three-dimensional physical space \cite{belbruno1993sun}, this according to the spatial stability definition provided in \cite{luo2014constructing}. Based on its dynamical behavior, a propagated trajectory is said to be:
i) \textit{weakly stable} (sub-set \Wset{i}) if the particle performs $i$ complete revolutions around the target neither escaping nor impacting with it or its moons;
ii) \textit{unstable} (sub-set \Xset{i}) if the particle escapes from the target before completing the $i$th revolution;
iii) \textit{target--crash} (sub-set \Kset{i}) if the particle impacts with the target before completing the $i$th revolution;
iv) \textit{moon--crash} (sub-set \Mset{i}) if the particle impacts with one of the target's moons before completing the $i$th revolution;
v) \textit{acrobatic} (sub-set \Dset{i}) if none of the previous conditions occurs within the integration time span.
Conditions ii)-v) apply after the particle performs $(i-\SI{1}{})$ revolutions around the target (see \Fig{fig:particle_stability}). The sub-sets are defined for $ i \in \mathbb{Z} \textbackslash \{0\} $, where the sign of $i$ informs on the propagation direction. When $ i > 0 $ ($ i < 0 $) the \gls*{IC} is propagated forward (backward) in time. A capture set is defined as $ \mathcal{C}_{-1}^{n} := \mathcal{W}_{n} \cap \mathcal{X}_{-1} $. Therefore, it is the intersection between the stable set in forward time \Wset{n} and the unstable set in backward time \Xset{-1} \cite{luo2014constructing}.

\Glspl{BCC} are time-varying manifolds supporting capture \cite{merisio2023engineering} obtained backward propagating \glspl{IC} belonging to a capture sets \Cset{n}, where $ n>0 $ is the number of revolutions after capture. They are defined in what follows. Firstly, a trajectory is defined as:
\begin{definition}
    Let $ (\mathbf{x}_{0},t_{0}) \in \mathbb{R}^{6} \times \mathbb{R} $ and $ {\bm{\varphi}}(\mathbf{x}_{0},t_{0};t) $ be the starting point and the solution at time $ t $, respectively, of the state-space representation $ \dot{\mathbf{x}} = \mathbf{f}(\mathbf{x},t) $ of the \gls*{EOM} in \Eq{eq:eom}. Then, a trajectory $\gamma$ is defined as $ \gamma(\mathbf{x}_{0},t_{0}) := \lbrace \bm{\varphi}(\mathbf{x}_{0},t_{0};t) \ \forall t \in \mathbb{R} \rbrace $.  Similarly, backward and forward legs $ \gamma_{b} $ and $ \gamma_{f} $, respectively, are defined as $ \gamma_{b} (\mathbf{x}_{0},t_{0}) := \lbrace \bm{\varphi}(\mathbf{x}_{0},t_{0};t) \ \forall t \in [t_{0}-\num{10} T_{\mars},\, t_{0}] \rbrace $ and $ \gamma_{f} (\mathbf{x}_{0},t_{0}) := \lbrace \bm{\varphi}(\mathbf{x}_{0},t_{0};t) \ \forall t \in [t_{0},\, t_{0}+\num{10} T_{\mars}] \rbrace $, where $ T_{\mars} = 2\pi \sqrt{a_{\mars}^{3} / \mu_{\Sun}} = \SI{687}{\days} $ is the revolution period of Mars, with $ a_{\mars} = \SI{2.279e8}{\km} $ and $ \mu_{\Sun} = \SI{1.327e11}{\kilo\meter\cubed\per\second\squared} $ the semi-major axis of the Sun--Mars system and the gravitational parameter of the Sun, respectively.
\end{definition}

Sets $ \Gamma_{\mathcal{W}_{n}} $, $ \Gamma_{\mathcal{X}_{-1}} $, and $ \Gamma_{\mathcal{C}_{-1}^{n}} $ of trajectories $ \gamma(\mathbf{x}_{0},t_{0}) $ whose \glspl{IC} $ (\mathbf{x}_{0},t_{0}) $ belong to weakly-stable set \Wset{n}, escape set \Xset{-1}, and capture set \Cset{n}, respectively, are $ \Gamma_{\mathcal{W}_{n}} = \lbrace \gamma(\mathbf{x}_{0},t_{0}) \ \forall \, (\mathbf{x}_{0},t_{0}) \in \mathcal{W}_{n} \rbrace $, $ \Gamma_{\mathcal{X}_{-1}} = \lbrace \gamma(\mathbf{x}_{0},t_{0}) \ \forall \, (\mathbf{x}_{0},t_{0}) \in \mathcal{X}_{-1} \rbrace $, and $ \Gamma_{\mathcal{C}_{-1}^{n}} = \lbrace \gamma(\mathbf{x}_{0},t_{0}) \ \forall \, (\mathbf{x}_{0},t_{0}) \in \mathcal{C}_{-1}^{n} \rbrace $. Similarly to a capture set \Cset{n}, a corridor is designated is defined as $ \mathcal{\check{B}}_{-1}^{n} = \lbrace \gamma_{b} (\mathbf{x}_{0},t_{0}) \ \forall \, (\mathbf{x}_{0},t_{0}) \in \mathcal{C}_{-1}^{n} \rbrace $. An exterior corridor \Ecorr{n} is a subset of a corridor \corridor{n} including pre-capture trajectories having heliocentric semi-major axis $ a_{\Sun} $ greater than the target body's one (\ie Mars, whose semi-major axis $ a_{t} = a_{\mars} = \SI{1.5237}{\AU} $). It is defined as $ \mathcal{\check{E}}_{-1}^{n} = \lbrace \gamma_{b} (\mathbf{x}_{0},t_{0}) \in \mathcal{\check{B}}_{-1}^{n}: a_{\Sun} \left( \bm{\varphi}(\mathbf{x}_{0},t_{0};t) \right) > a_{t} \ \forall t \in [t_{0} - \num{10} T_{\mars},\, \hat{t}] \rbrace $ where $ \hat{t} < t_{0} $ is a certain time before capture epoch $ t_{0} $ when the escape (or pre-capture) leg ends in backward time. Contrarily, an interior corridor \Icorr{n} is the subset of a corridor \corridor{n} including all trajectories having semi-major axis smaller than the central body's one (\ie Mars). It is defined as $ \mathcal{\check{I}}_{-1}^{n} = \lbrace \gamma_{b} (\mathbf{x}_{0},t_{0}) \in \mathcal{\check{B}}_{-1}^{n}: a_{\Sun} \left( \bm{\varphi}(\mathbf{x}_{0},t_{0};t) \right) < a_{t} \ \forall t \in [t_{0} - \num{10} T_{\mars},\, \hat{t}] \rbrace $. Consequently, $ \mathcal{\check{B}}_{-1}^{n} = \mathcal{\check{E}}_{-1}^{n} \cup \mathcal{\check{I}}_{-1}^{n} $.

\begin{figure}[t]
    \centering
    \includegraphics[width=\linewidth,clip,trim={0 0 0 0}]{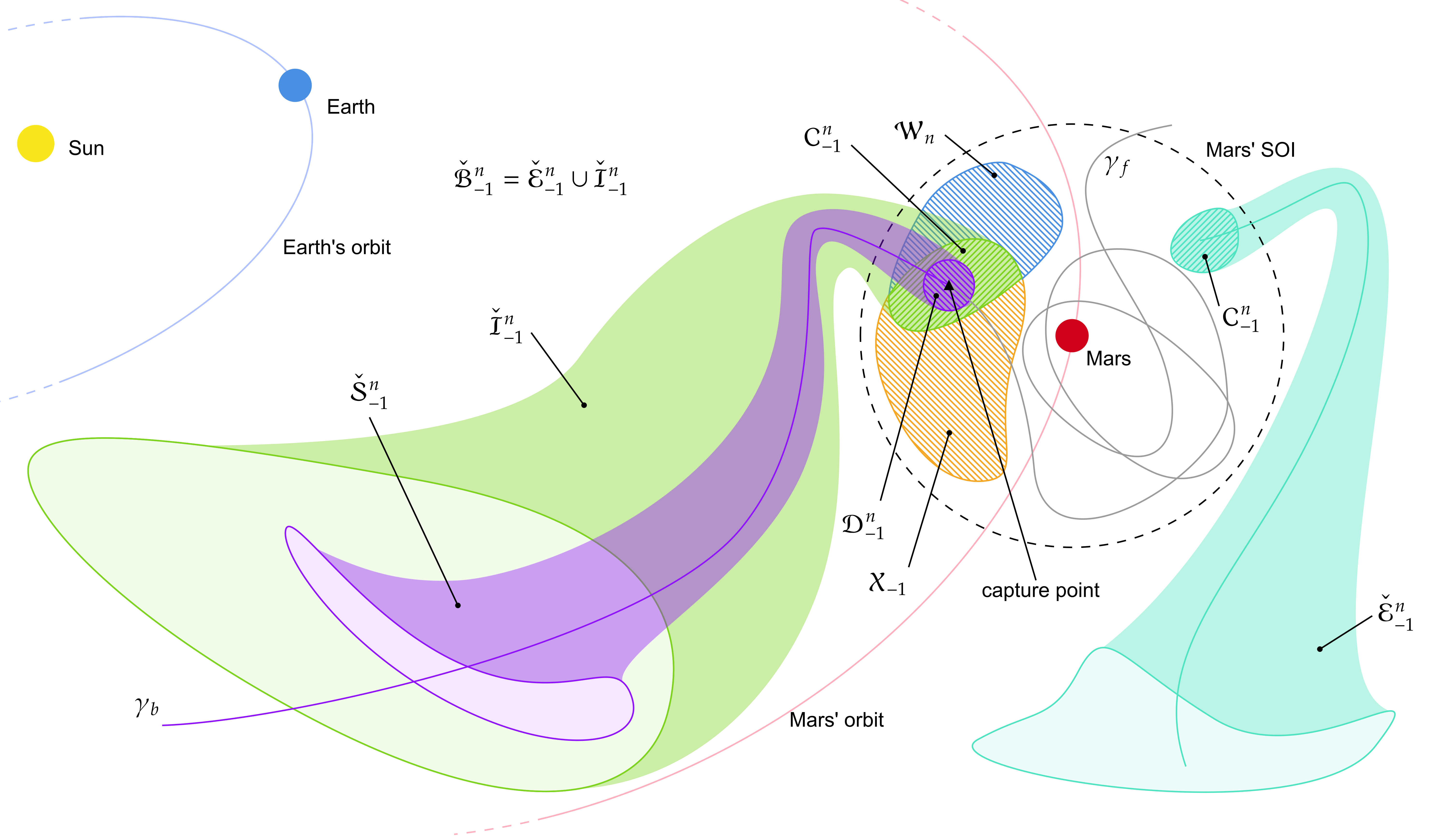}
    \caption{Illustration of BCC definitions.}
    \label{fig:bcc_concept}
\end{figure}

In this work, the interior corridor is of interest because it extends between Mars and Earth's orbits. A subcorridor \Scorr{n}, a generic subset of a corridor \corridor{n}, is defined as $ \mathcal{\check{S}}_{-1}^{n} = \lbrace \gamma_{b} (\mathbf{x}_{0},t_{0}) \ \forall \, (\mathbf{x}_{0},t_{0}) \in \mathcal{D}_{-1}^{n} \rbrace $, where the generic domain is $ \mathcal{D}_{-1}^{n} = \lbrace \mathbf{x}_{0}: \mathbf{x}_{0} \in \mathcal{C}_{-1}^{n} \wedge \mathbf{g}(\mathbf{x}_{0}) \leq \mathbf{0} \wedge \mathbf{h}(\mathbf{x}_{0}) = \mathbf{0} \rbrace $, with $ \mathbf{g}(\mathbf{x}_{0}) $ and $ \mathbf{h}(\mathbf{x}_{0}) $ being two sets of $ m \geq 0 $ inequality constraints and $ n \geq 0 $ equality constraints, respectively, with $ m $ and $ p $ finite. Finally, the envelope \Envcorr{n} of a subcorridor \Scorr{n} is constructed backward propagating the subcorridor domain border $ \partial \mathcal{D}_{-1}^{n} $. Therefore, it is defined as $ \partial \mathcal{\check{S}}_{-1}^{n} = \lbrace \gamma_{b} (\mathbf{x}_{0},t_{0}) \ \forall \, (\mathbf{x}_{0},t_{0}) \in \partial \mathcal{D}_{-1}^{n} \rbrace $. An illustration of prior definitions is proposed in \Fig{fig:bcc_concept}.

\subsection{\Acrlong*{PCE}} 

\Gls*{PCE} is an uncertain quantification method able to provide an efficient mean for long term propagation of non-Gaussian distributions. \gls*{PCE} approximates the stochastic solution of the governing dynamics as a weighted sum of multivariate spectral polynomials, function of the input random variables \cite{jones2013nonlinear}
\begin{equation}
\label{eq:PCE_exp}
    \bm{x}(t,\bm{\xi}) \approx \hat{\bm{x}}(t,\bm{\xi})=\sum_{\bm{\alpha} \in \Lambda_{p,d}} \bm{c}_{\bm{\alpha}}(t)\psi_{\bm{\alpha}}(\bm{\xi})
\end{equation}
where $\bm{\xi} \in \mathbb{R}^d$ is the random input vector, $\bm{c}_{\bm{\alpha}}(t)$ is the vector of \glspl{PCC}, and $\Lambda_{p,d}$ is the set of multi-indices of size $d$ and order $p$, having so a total dimension equal to \cite{jones2013nonlinear}
\begin{equation}
	\label{eq:PCE_P2}
	P=|\Lambda_{p,d}|=\dfrac{(p+d)!}{p!\,d!}.
\end{equation}
Namely, $\bm{c}_{\bm{\alpha}}(t) \in \mathbb{R}^n$, where $n$ is the number of quantities of interest. If considering the full state propagation, $\bm{x}$ represents both the position and velocity and, hence, $n=6$.

The basis functions $\psi_{\bm{\alpha}}(\bm{\xi})$ are multidimensional spectral polynomials, orthonormal with respect to the joint probability measure of the vector $\bm{\xi}$. Thus, the basis functions choice depends only on the properties of the input variables. For example, Hermite and Legendre polynomials are the basis for the Gaussian and uniform distributions, respectively \cite{xiu2002wiener}.

Generating a \gls*{PCE} means computing the \glspl*{PCC} by projecting the exact solution onto each basis function $\psi_{\bm{\alpha}}(\bm{\xi})$ \cite{jones2013nonlinear}
\begin{equation}
    \label{eq:PCE_exact}
    \bm{c_\alpha}(t)=\mathit{E}[\bm{x}(t,\cdot)\psi_{\bm{\alpha}}(\cdot)]= \int_{\Gamma^d} \bm{x}(t,\bm{\xi}) \psi_{\bm{\alpha}}(\bm{\xi})\rho(\bm{\xi}) d\bm{\xi}
\end{equation}
where $\Gamma^d$ is the $d$-dimensional hypercube where the random input variables are defined. Once the \glspl{PCC} are computed, the system state associated to a specific sample within the capture subset can be retrieved at time $t$ effortlessly using Eq.~\eqref{eq:PCE_exp}. The \glspl{PCC} estimation methods fall into two categories: intrusive and non-intrusive. While the first requires laborious modifications in the governing equations, the latter treat the dynamics as a black box, thus they are better suited for  problems with high-fidelity complex dynamics \cite{giordano2021analysis}.

\section{Methodology} \label{sec:method}

\subsection{Problem statement}

The goal of the \gls*{BCC} synthesis is to produce a numerical approximation of a subcorridor. The approximation is later made available to the autonomous guidance and control unit implemented onboard the limited-capability spacecraft. Ideally, the evaluation of the synthetic subcorridor shall be fast and inexpensive for spacecraft having limited onboard resources. In mathematical terms, the \textit{general subcorridor synthesis problem} can be thus stated as follows \cite{merisio2023engineering}:

\begin{problem}
    Numerically synthesize the subcorridor $ \mathcal{\check{S}}_{-1}^{n} = \lbrace \gamma_{b} (\mathbf{x}_{0},t_{0}) \ \forall \, (\mathbf{x}_{0},t_{0}) \in \mathcal{D}_{-1}^{n} \rbrace $ as a function $ \mathbf{x} = \bm{\psi} (\mathbf{p}) $ of some parameters $ \mathbf{p} $ such that, given the parameters $ \mathbf{p}^{*} $, the state $ \mathbf{x}^{*} = \bm{\psi} (\mathbf{p}^{*}) $ is retrieved. In particular, the state $ \mathbf{x}^{*} $ must be targeted by the spacecraft to be temporarily captured by the central body at time epoch $ t_{0} $, so performing at least $ n $ revolutions about it.
\end{problem}

The choice of the parameters $ \mathbf{p} $ to be selected as support of the subcorridor is of paramount importance, since they define the input for the synthesis. The aim is to find a set of coordinates for which the regular capture sub-region is sufficiently large. Indeed, the wider the capture subset considered, the bigger the region to target in the interplanetary leg. For the purpose of this work, capture sets at Mars are built following the methodology depicted in \cite{luo2020role}. According to the methodology the initial computational grid of \glspl{IC} is bidimensional, thus it is reasonable to use two parameters to define the subcorridor. The most significant results were achieved in Cartesian and Keplerian coordinates. To be compliant with the approach discussed in \cite{merisio2022algorithm}, two components of the Cartesian coordinates (namely, $x$ and $y$) have been chosen to properly represent the capture set. Additionally, since the \gls*{BCC} is designed propagating a conveniently selected capture subset, which ensures ideal post-capture behavior \cite{merisio2022algorithm, morelli2022convex}, the probability distribution of initial condition in the capture subset can be considered uniform, i.e., each condition within the subset boundaries leads to capture. This assumption paves the way to the use of \gls*{PCE} as an efficient synthetization method, since initial \gls*{BC} conditions can be treated as stochastic variables. Hence, a \textit{PCE-based subcorridor synthesis problem} can be stated as follows:

\begin{problem}
	Find the \acrlongpl{PCC} $\bm{c}_{\bm{\alpha}}(t)$, such that $\bm{x}\left(t,\bm{\xi}\right)=\sum_{\bm{\alpha} \in \Lambda_{p,d}} \bm{c}_{\bm{\alpha}}(t)\psi_{\bm{\alpha}}(\bm{\xi})$, with $\bm{\xi}=(x_0, y_0)$ numerically synthesize the subcorridor $ \mathcal{\check{S}}_{-1}^{n} = \lbrace \gamma_{b} (\mathbf{x}_{0},t_{0}) \ \forall \, (\mathbf{x}_{0},t_{0}) \in \mathcal{D}_{-1}^{n} \rbrace $.
\end{problem}

\subsection{\Acrlong*{PCE} application}

Among the different non-intrusive \gls{PCE} techniques, given the low number of input parameters (i.e., $x$ and $y$ coordinates of the capture set \glspl{IC}), the pseudospectral collocation approach with full tensor grid \cite{jones2013nonlinear} is selected. As a matter of fact, bi-dimensional quadrature schemes (i.e., with $d=2$) suffer in a limited way of the curse of dimensionality, while pseudospectral collocation approaches guarantee a simple and accurate method for the \glspl{PCC} computation. In this case, \glspl{PCC} can be computed by solving the stochastic integral \cite{jones2013nonlinear}

\begin{equation}
	\bm{c}_{\bm{\alpha}}(t)\asymp\mathcal{Q}\left[\bm{x}(t,\cdot)\psi_{\bm{\alpha}}(\cdot)\right]=\sum_{i=1}^d\sum_{q_i=1}^{m_i}\bm{x}\left(t,\bm{\xi}_i^{q_i}\right)\psi_{\bm{\alpha}}(\bm{\xi}_i^{q_i})\omega_{q_i}
	\label{eq:StocInt}
\end{equation}

where $\mathcal{Q}$ is the quadrature integration, $\{\boldsymbol{\xi}_q\}$ is the set of quadrature nodes and $\{\omega_q\}$ are the quadrature weights. Several quadrature rules are available. Gaussian quadrature is exploited in this work due to their high degree of accuracy \cite{xiu2010numerical}. Consequently, nodes $\boldsymbol{\xi}_q$ and corresponding weights $\omega_q$ are selected depending on the orthogonal polynomials associated to the probability density function $\tilde{\rho}(\xi_i)$ related to each random input $\xi_i$. In this case, since the input $\xi_i=(x_0^i, y_0^i)$ is described as a uniform random variable, the nodes $\{\xi_i^{q_i}\}$ and the weights $\{\omega_{q_i}\}$ with $q_i=1,\dots,m_i$ are the zeros of the Legendre polynomial of order $m_i$ and its quadrature weights.

\section{Results} \label{sec:results}

\subsection{Capture subset dimension} \label{sec:cap_set_dim}

\begin{figure}
	\centering
	\includegraphics[scale=0.95]{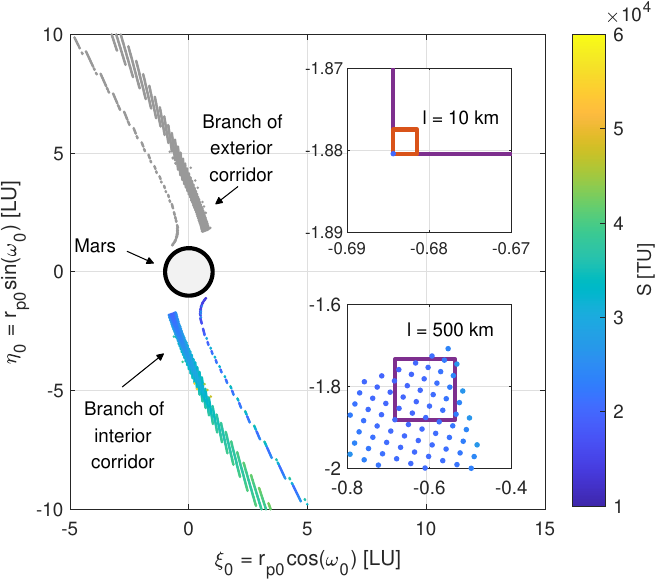}
	\caption{Capture set \Cset{2} at $ t_{0} = $ December 9, 2023 at 12:00:00 (UTC). Branches developing exterior BBCs are colored in gray. Regularity index \cite{deitos2018survey} of ICs belonging to branches developing interior BCCs. Nondimensional coordinates on the orbital plane $ i_{0} = \Omega_{0} = \num{0.2} \pi \, \si{\radian} $ defined in the Mars-centered RTN@$ t_{0} $ frame. Mars is the gray circle with black surround. In the magnifications, details of chosen capture subsets.}
	\label{fig:capture_set}
\end{figure}

Without any loss of generality, the analysis has been conducted on the capture set depicted in \Fig{fig:capture_set}. First, a squared capture subset should be identified according to the generic domain $ \mathcal{D}_{-1}^{n} $ definition provided in \Sec{sec:bcc}. This is done by identifying one point, which is taken to be the bottom-left vertex of the subdomain, and a square side length $ l $. The bottom-left vertex of the domain is associated to the following $ (r_0, \omega_0) $ pair : $ r_0 = R_t + \SI{3400}{\kilo\meter} $ and $ \omega_0 = \SI{250}{\deg} $, where $ R_t $ is the target (\ie Mars) mean radius. The square side $l$ range is selected to span the cluster width completely. The lower limit is set to \SI{10}{\kilo\meter}, while the upper one to \SI{500}{\kilo\meter}, yielding $ l \in [\num{10}, \num{500}] \, \si{\kilo\meter} $, as shown in \Fig{fig:capture_set}.

\begin{figure}
    \centering
    \subfloat[Position error $ \varepsilon_r $.]{\includegraphics[width=0.5\textwidth]{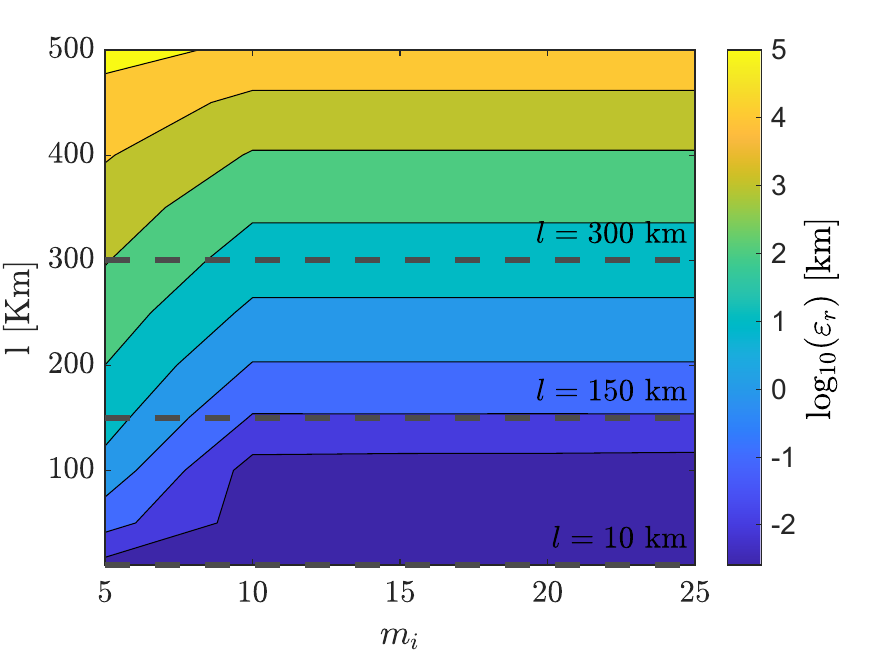}}
    \subfloat[Velocity error $ \varepsilon_v $.]{\includegraphics[width=0.5\textwidth]{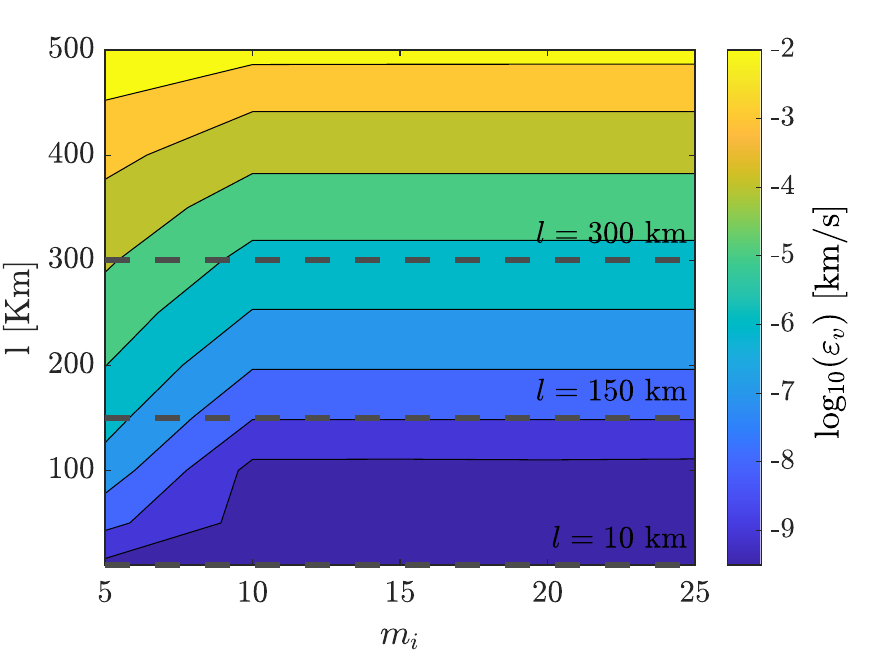}}
    \caption{Approximation error at $ t_f $ as a function of quadrature nodes number $ m_i $ and square side length $ l $. Polynomial basis order fixed to $ p = 6 $. Specific subsets having constant $l$ highlighted with dashed horizontal lines.}
    \label{fig:contour_fix-p}
\end{figure}

To verify the applicability of \gls*{PCE} to the subcorridor synthesis, the influence of the subset dimensions on the expansion parameters (\ie $ m_i $ and $ p $) is investigated. Firstly, the correlation between the quadrature nodes number $ m_i $ and the side length $ l $ is assessed. For this analysis, the polynomial basis order is kept constant to $ p = 6 $. The range [5, 25] is assumed for $ m_i $. The outcome is displayed in \Fig{fig:contour_fix-p}. The accuracy is evaluated at the beginning of the pre-capture trajectory (\ie epoch $ t_f = t_0 - \SI{400}{\days} $). For $ 5 \leq m_i \leq 10 $, the level curves show a ramp-like behavior. Intuitively, the accuracy increases with the density of quadrature nodes. For instance, consider a side length $ l = \SI{150}{\kilo\meter} $, by progressively increasing $ m_i $ from 5 to 10, it improves by three digits, both in position and velocity. However, for $ m_i > 10 $, a plateau is reached. This is because theoretical results suggests that, for an increasing number of nodes, the estimation accuracy does not improve significantly once the estimation is exact. With exact estimation, it is intended the Gaussian quadrature exactness for polynomial of degree $ p \leq (2 m_i - 1) $ \cite{jones2013nonlinear}. On the other hand, the method seems to fail for $ l = \SI{300}{\kilo\meter} $. The error in position is stuck in the order of \si{\kilo\meter} and accuracy stops improving with quadrature nodes number.

\begin{figure}
    \centering
    \subfloat[Position error $ \varepsilon_r $.]{\includegraphics[width=0.5\textwidth]{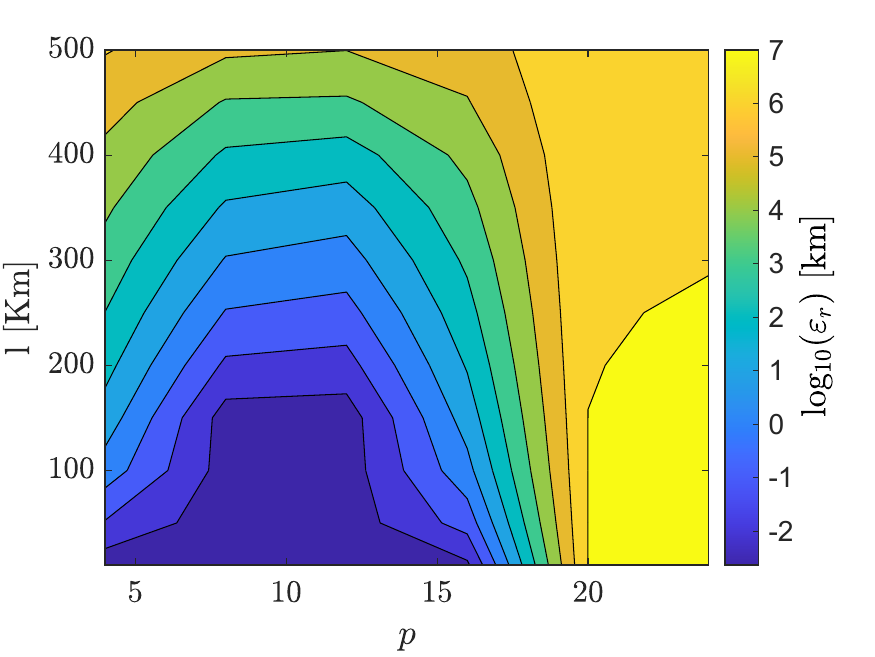}}
    \subfloat[Velocity error $ \varepsilon_v $.]{\includegraphics[width=0.5\textwidth]{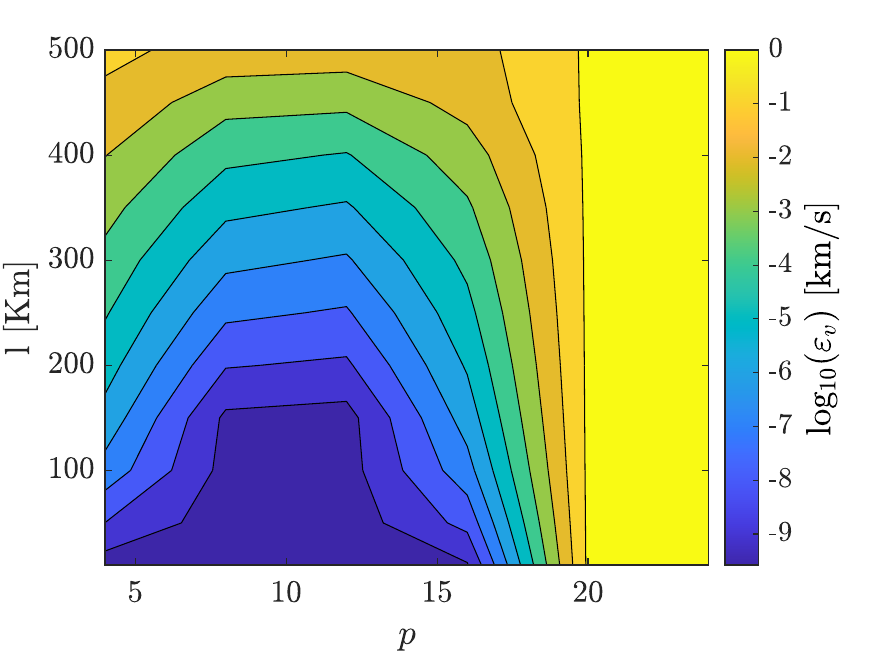}}
    \caption{Approximation error at $ t_f $ as a function of polynomial basis order $p$ and square side length $ l $. Quadrature nodes number fixed to $ m_i = 10 $.}
    \label{fig:contour_fix-m}
\end{figure}

Next, the correlation between the polynomial basis order $ p $ and the side length $ l $ is investigated. For this analysis, the quadrature nodes number is kept constant to $ m_i = 10 $. Results are shown in \Fig{fig:contour_fix-m}. Remarkably, the method accuracy is greatly affected by the polynomial basis order. Indeed, the method accuracy improves by progressively increasing $ p $. For $ l = \SI{100}{\kilo\meter} $, the approximation errors decrease by one order of magnitude when $ p $ increases from 6 to 10. Nevertheless, accuracy is lost when increasing indefinitely the polynomial basis order. As clearly visible on the right of plots in \Fig{fig:contour_fix-m}, increasing $ p $ results in being a disadvantage for the method accuracy. This behavior may be justified considering that accuracy of Gaussian quadrature rule is exact for polynomials of degree $ p \leq (2 m_i - 1) $ \cite{jones2013nonlinear}. The latter translates into the necessity of rising the quadrature nodes density as $ p $ increases. From \Fig{fig:contour_fix-m}, it is clear how for large $ p $ values, the quadrature rule fails in estimating accurately the integral in \Eq{eq:StocInt}, thereby growing the approximation error.

\subsection{Simulation parameters}

\begin{figure}
    \centering
    \subfloat[Position error $ \varepsilon_r $.]{\includegraphics[width=0.5\textwidth]{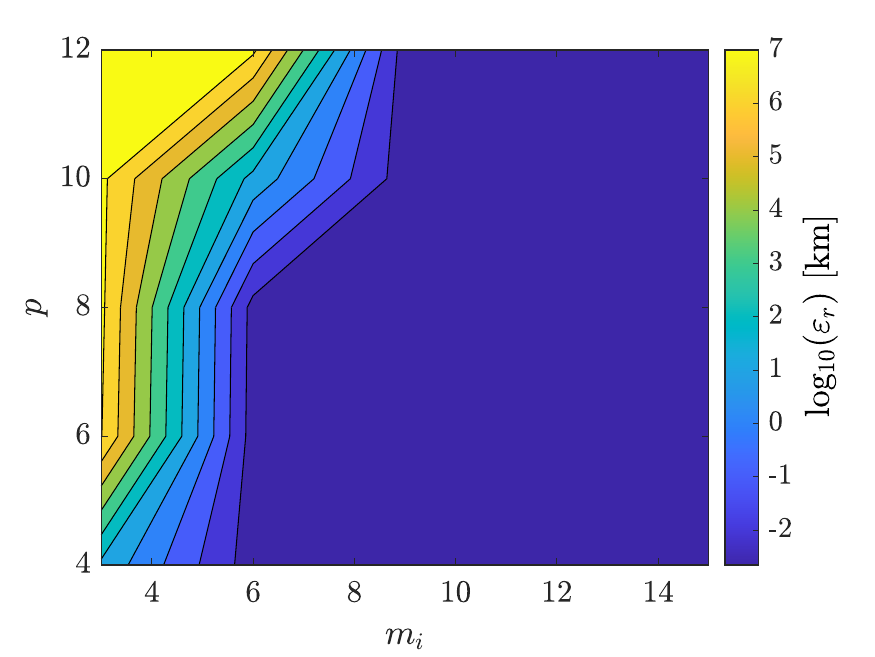}}
    \subfloat[Velocity error $ \varepsilon_v $.]{\includegraphics[width=0.5\textwidth]{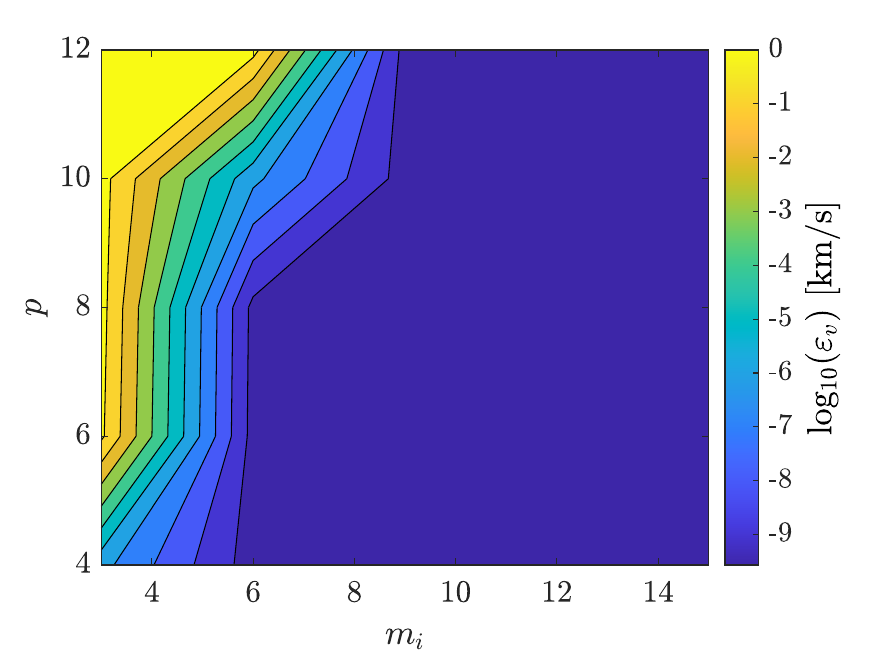}}
    \caption{Approximation error at $ t_f $ as a function of quadrature nodes number $ m_i $ and polynomial basis order $ p $. Square side length fixed to $ l = \SI{10}{\kilo\meter} $.}
    \label{fig:contour_fix-l}
\end{figure}

The correlation between $ m_i $ and $ p $ is now investigated by fixing the square side length of the chosen capture subset to $ l = \SI{10}{\kilo\meter} $. In this analysis, the quadrature nodes number $ m_i $ and the polynomial basis function $ p $ are let vary in ranges [3, 15] and [4, 12], respectively. The outcome is proposed in \Fig{fig:contour_fix-l}. As already noted, the method accuracy is almost unchanged for large quadrature nodes numbers. Furthermore, for $ m_i < 9 $, accuracy is lost as the polynomial basis order $ p $ increases. The latter confirms what was previously discussed in \Sec{sec:cap_set_dim}. 

\subsection{Polynomial chaos coefficients distribution}

\begin{figure}
\centering
    \subfloat[$x$ component.\label{fig:PCCx}]{\includegraphics[width=0.48\textwidth]{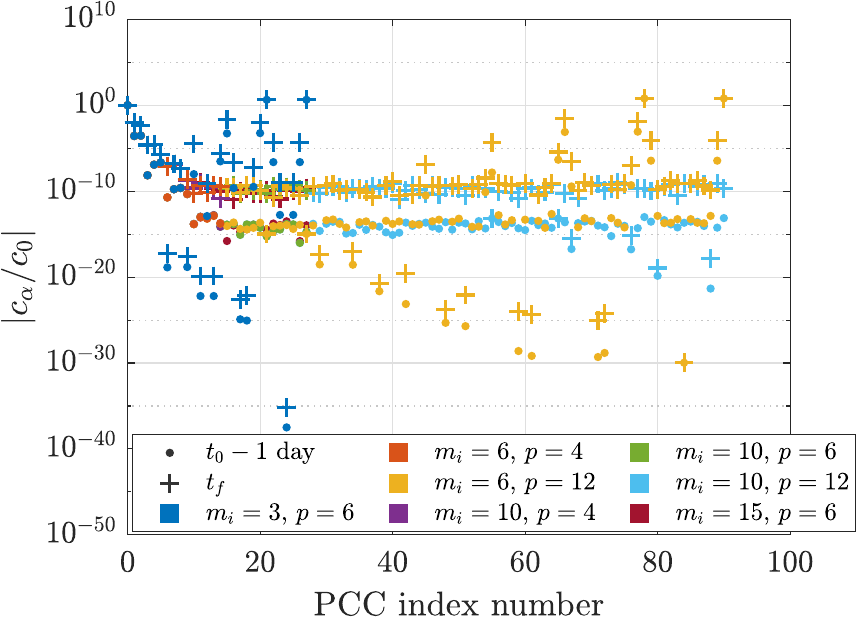}}
    \quad
    \subfloat[$y$ component.\label{fig:PCCy}]{\includegraphics[width=0.48\textwidth]{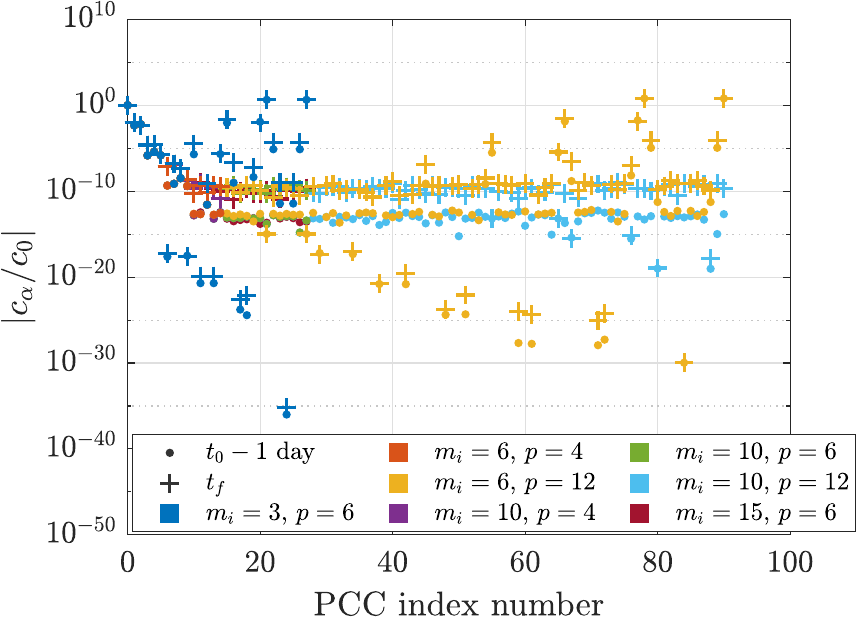}}

    \subfloat[$v_x$ component.\label{fig:PCCvx}]{\includegraphics[width=0.48\textwidth]{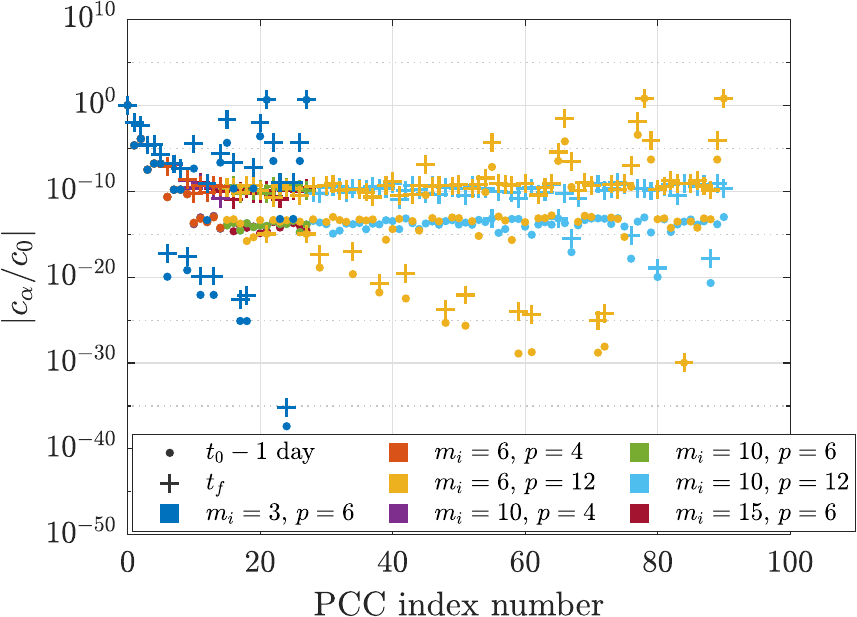}}
    \quad
    \subfloat[$v_z$ component.\label{fig:PCCvz}]{\includegraphics[width=0.48\textwidth]{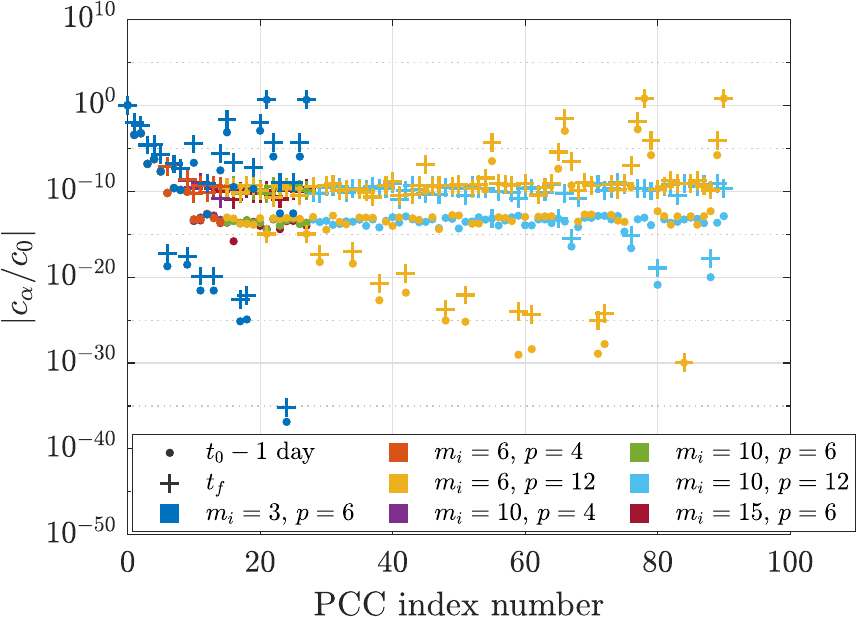}}
    
    \caption{Normalized PCE coefficients distributions for 4 out of the 6 state variables. Dots and crosses indicate coefficients distribution \SI{1}{} and \SI{400}{\days} before capture, respectively. Note that \SI{400}{\days} before capture corresponds to the final backward propagation epoch $t_f$. Several $ (m_i, p) $ pairs shown.}
    \label{fig:PCCplot}
\end{figure}

The \glspl*{PCC} distribution for several $ (m_i, p)$ pairs, and with $ l = \SI{10}{\kilo\meter} $ are shown in \Fig{fig:PCCplot}. In the plot, dots denote distributions \SI{1}{\day} before capture, while crosses the distributions \SI{400}{\days} before capture (corresponding to epoch $ t_f $). Coefficients are normalized with respect to coefficient $ c_0 $. The higher the decay rate, the more accurate is the expansion result. This is because higher order terms become less relevant \cite{jones2013nonlinear}.

Results show that for the (10, 6) pair convergence is achieved at both epochs. Namely, at $ t_0 - \SI{1}{\day} $ the distribution converges to within machine precision ($\sim \! 10^{-15}$). Differently, the normalized coefficients converge at a lower pace to a higher asymptotic value ($\sim \! 10^{-10})$ at $ t_f $. This is consistent with results presented in \cite{jones2013nonlinear}, according to which the convergence value can be correlated to the digit precision in the state estimate. Indeed, numerical errors increase with the propagation time, leading to lower digit precision. Nonetheless, the digit precision is very high, both in nondimensional position and velocity, therefore results are satisfactory.

Varying the quadrature nodes number, the expansion fails in achieving convergence for $ m_i = 3 $. According to results in \Fig{fig:contour_fix-l}, $ m_i > 4 $ is recommended. Contrarily, no significant variations are observed between the (10, 6) pair and the case with $ m_i = 15 $. This is consistent with results in \Figs{fig:contour_fix-p} and~\ref{fig:contour_fix-l}. Differently, by letting $ p $ vary, convergence is assessed for (10, 4) and (10, 12) pairs. However, all high-order polynomials contribute marginally to the expansion accuracy for the (10, 12) pair. The latter representation is usually referred to as sparse \cite{jones2013nonlinear}. Thus, only the most significant expansion coefficients are retained when $ p = 4 $, thereby providing a good estimate of the system state. Finally, the top-right region of \Fig{fig:contour_fix-l} is investigated. Convergence is achieved for the (6, 4) pair. Differently, the (6, 12) pair fails to converge since the quadrature rule is unsuccessful in accurately estimating the \glspl*{PCC} for $ p > (2 m_i - 1) $ \cite{jones2013nonlinear}.

\subsection{Validation}

The \gls*{PCE} technique is validated against \gls*{MC} analysis. In the following, the convergence analysis focuses on a specific case study, which develops the corridor upon the same capture subset with $ l = \SI{10}{\kilo\meter} $ described in \Sec{sec:cap_set_dim}. The selected simulation parameters are $ m_i = 10 $ and $ p = 6 $.

\begin{figure}
    \centering
    \subfloat[Position error.]{\includegraphics[width=0.46\textwidth]{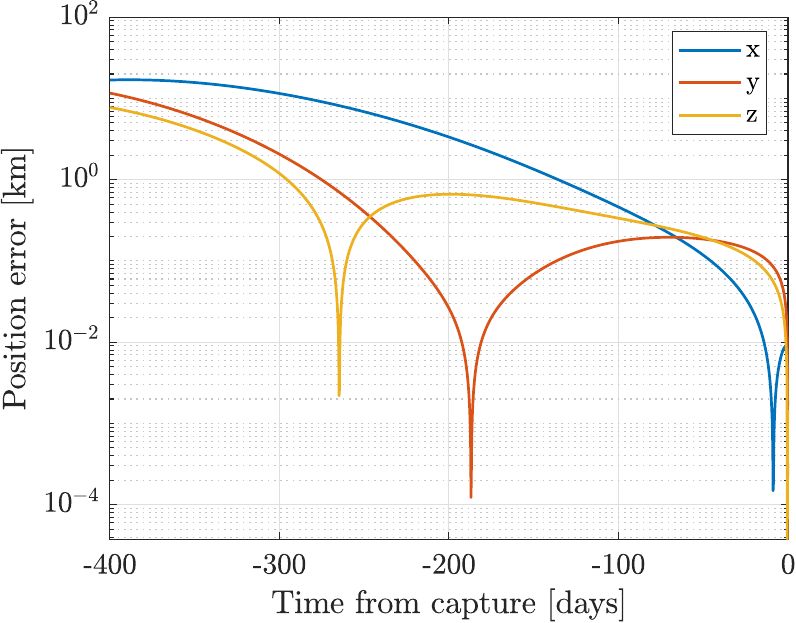}}
    \qquad
    \subfloat[Velocity error.]{\includegraphics[width=0.46\textwidth]{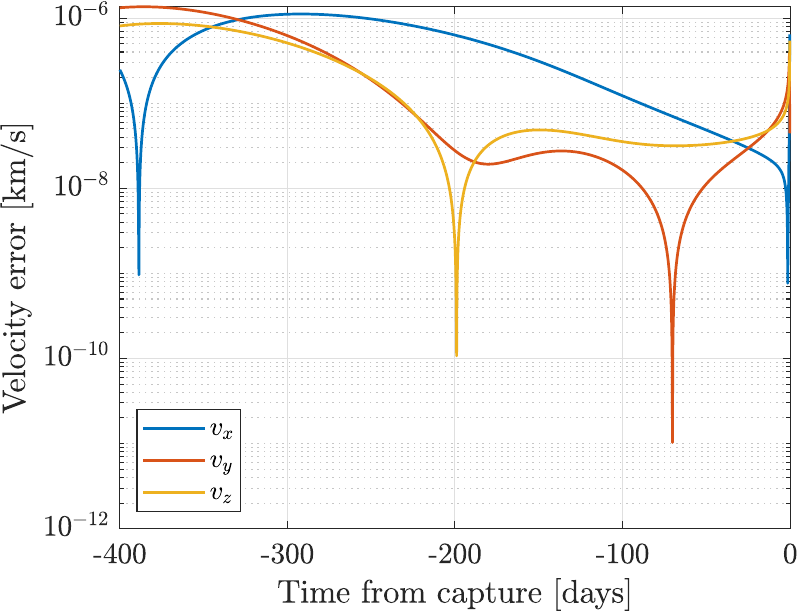}}
    \caption{Method accuracy in estimating the mean. Evolution of first PCC compared to estimated mean drawn from capture subset with LHS technique and computed from $ 10^2 $ samples.}
    \label{fig:abs_mean_err_time}
\end{figure}

The convergence analysis is evaluated at fixed time epoch. Therefore, the epoch is suitably selected to assess the \gls*{PCE} validity in approximating the pre-capture trajectory. To this aim, the \gls*{PCE} and \gls*{MC} approaches are compared considering their evolution in time. Namely, the \gls*{PCE} estimated mean is compared with the mean estimate computed from $ 10^2 $ samples. Samples are drawn with the \gls*{LHS} technique. In \Fig{fig:abs_mean_err_time}, the approximation error is evaluated as $ \bm{e}(t) = |\bm{c}_{0}(t) - \bm{\mu}(t)| $, where $ \bm{c}_{0}(t) $ is the first \gls*{PCC}, being the mean according to the \gls{PCE}, and $ \bm{\mu}(t) $ is the mean value retrieved with the \gls*{MC} analysis. From results, it is observed that the error increases with the propagation time (from $ t_0 $ to $ t_f = \SI{-400}{\days} $). It comes naturally to assess the method convergence at the final epoch $ t_f $, which is associated with the largest approximation error. 

\begin{figure}
\centering
    \subfloat[Mean $ \mu_x $.]{\includegraphics[width=0.48\textwidth]{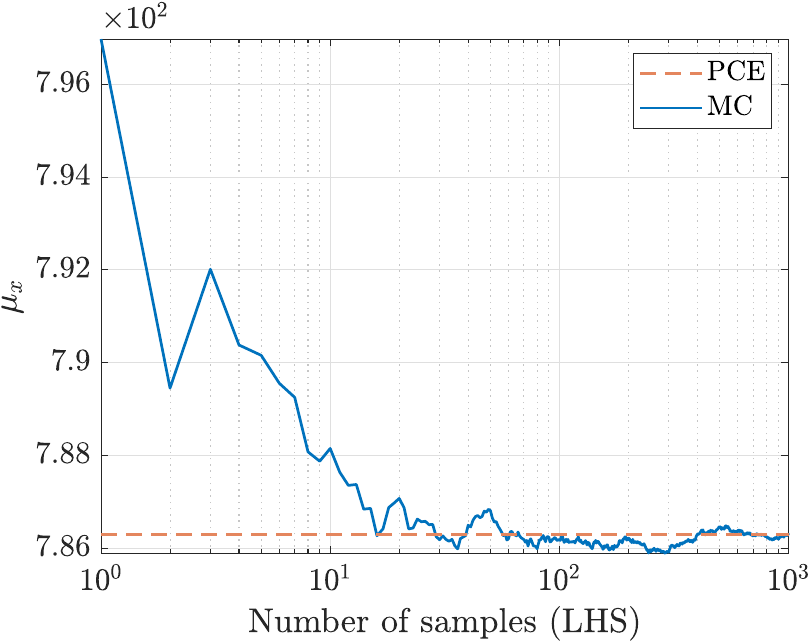}}
    \quad
    \subfloat[Mean $ \mu_{v_y} $.]{\includegraphics[width=0.48\textwidth]{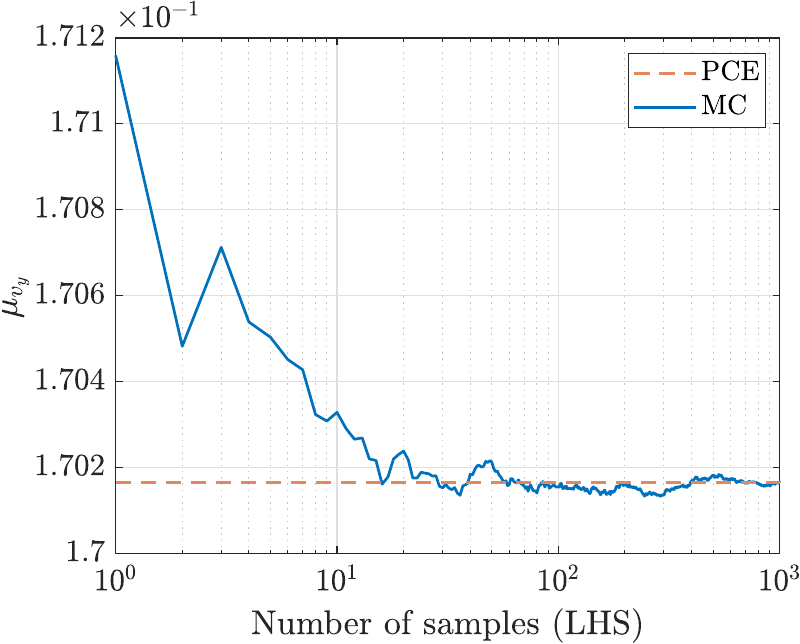}}
    
    \subfloat[Standard deviation $ \sigma_x $.]{\includegraphics[width=0.48\textwidth]{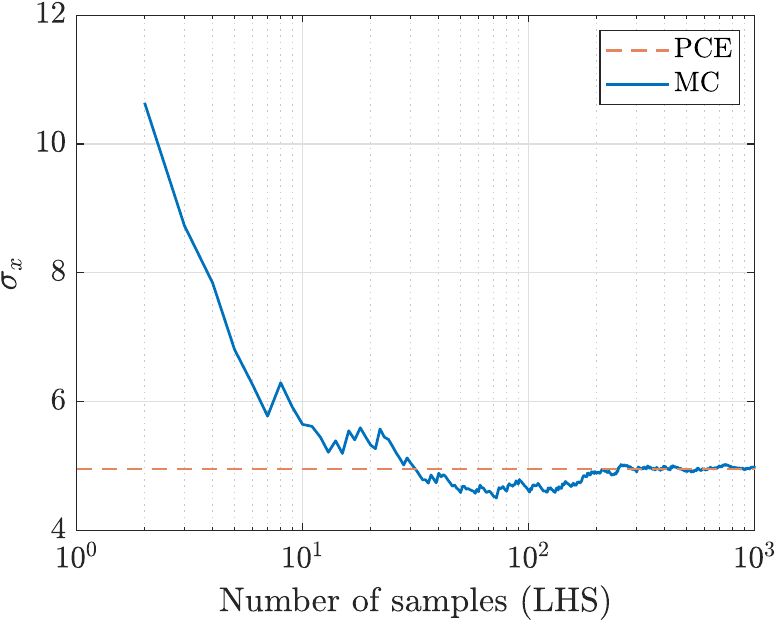}}
    \quad
    \subfloat[Standard deviation $ \sigma_{v_y} $.]{\includegraphics[width=0.48\textwidth]{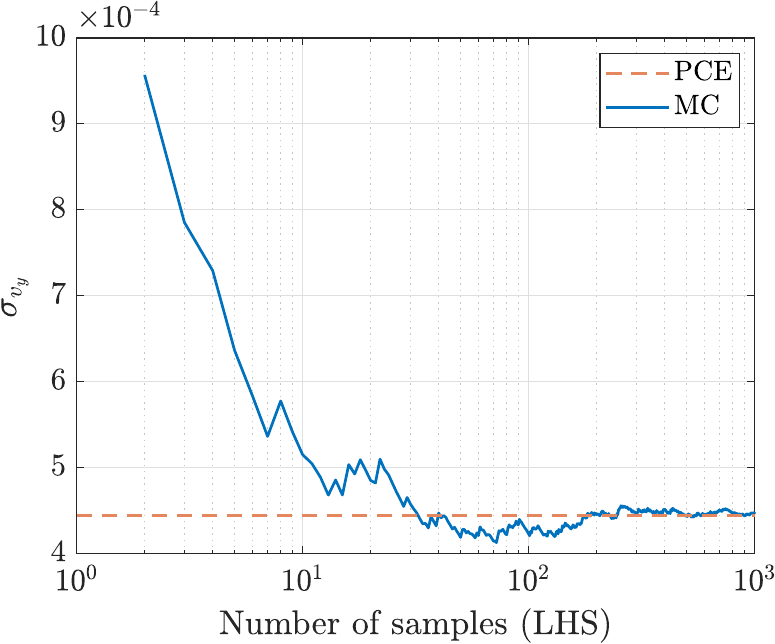}}
    \caption{Estimated mean and standard deviation at $ t_f = t_0 - \SI{400}{\days} $ for 2 out of the 6 state variables. Results from MC simulation (solid blu line) and PCE (dashed red line) as a function of samples number. Nondimensional quantities (see \Tab{tab:units}) in the Mars-centered J2000 frame.}
    \label{fig:conv}
\end{figure}

The convergence is evaluated on the mean and standard deviation. \Fig{fig:conv} compares the results for the \gls*{MC} simulation and the \gls*{PCE} technique as a function of the samples $ n $. Convergence errors are computed as
\begin{equation}
    \varepsilon_{\mu_r} = \max_{i = x, y, z} \left| \dfrac{\mu^{\mathrm{MC}}_i(t_f) - \mu^{\mathrm{PCE}}_i(t_f)}{\mu^{\mathrm{PCE}}_i(t_f)} \right|,
    \qquad\
    \varepsilon_{\mu_v} = \max_{i = v_x, v_y, v_z} \left| \dfrac{\mu^{\mathrm{MC}}_i(t_f) - \mu^{\mathrm{PCE}}_i(t_f)}{\mu^{\mathrm{PCE}}_i(t_f)} \right|,
\end{equation}
\begin{equation}
    \varepsilon_{\sigma_r} = \max_{i = x, y, z} \left| \dfrac{\sigma^{\mathrm{MC}}_i(t_f) - \sigma^{\mathrm{PCE}}_i(t_f)}{\sigma^{\mathrm{PCE}}_i(t_f)} \right|,
    \qquad\
    \varepsilon_{\sigma_v} = \max_{i = v_x, v_y, v_z} \left| \dfrac{\sigma^{\mathrm{MC}}_i(t_f) - \sigma^{\mathrm{PCE}}_i(t_f)}{\sigma^{\mathrm{PCE}}_i(t_f)} \right|.
\end{equation}
Referring to \Fig{fig:conv}, the mean and standard deviation convergence errors for $ n = 10^3 $ are $ \varepsilon_{\mu_r} = \num{5.682e-8} $, $ \varepsilon_{\mu_v} = \num{2.924e-12} $, $ \varepsilon_{\sigma_r} = \num{4.943e-5} $, and $ \varepsilon_{\sigma_v} = \num{5.4235e-7} $. Eventually, the \gls*{MC} analysis converges to the reference solution with a large degree of accuracy. Consistently with theoretical results predicted in the literature, the standard deviation converges at a lower rate with respect to the mean \cite{xiu2010numerical}. Therefore, the \gls*{PCE} technique is validated against the \gls*{MC} approach for the problem at hand. Remarkably, the \gls*{MC} approach with \gls*{LHS} requires the propagation of $10^3$ samples, whereas only $ m_i^d = 10^2 $ propagations are required by the \gls*{PCE} method when applied to this case study.  

\section{Conclusion} \label{sec:conclusion}

In this paper, a procedure to accurately and inexpensively synthesize \acrlongpl*{BCC} exploiting the \acrlong*{PCE} technique is discussed and validated against \acrlong*{MC} simulations. Results prove the convergence of the method, assess the feasibility of \acrlong*{BCC} numerical synthesis, and highlight its convenience in terms of computational efficiency. Remarkably, as the capture subset dimension increases, the method accuracy is preserved by properly tuning the quadrature nodes number and the polynomial basis order. For constant polynomial basis order, the method accuracy improves as the number of quadrature nodes increases up to the point a plateau is reached. Denser quadrature nodes imply higher computational costs, reducing the computational efficiency and making the \acrlong*{BCC} construction more expensive. On the other hand, for fixed quadrature nodes number, the method accuracy does not improve by increasing the polynomial basis order. Indeed, the method accuracy decreases because the quadrature nodes are insufficient, thereby poorly estimating high-order polynomials. Overall, results show that a convenient combination of quadrature nodes number and polynomial basis order improves the accuracy of the method at limited computational costs.

\section*{Funding Sources}
The authors would like to acknowledge the \gls*{ERC} since part of this work has received funding from the \gls*{ERC} under the European Union’s Horizon 2020 research and innovation programme (Grant Agreement No.\,864697).

\bibliography{references.bib}

\end{document}